\documentclass[12pt,preprint]{aastex}

\newcommand{\mm}{\hat{M}}
\newcommand{\y}{\tilde{x}}
\newcommand{\z}{\tilde{z}}

\shorttitle{ACCELERATION OF JETS}
\shortauthors{TOMIMATSU and TAKAHASHI}

\begin{document}


\title{RELATIVISTIC ACCELERATION OF  \\
    MAGNETICALLY DRIVEN JETS}

\author{AKIRA TOMIMATSU}
\affil{Department of Physics, University of Nagoya, Nagoya 464-8602, Japan}
\email{atomi@allegro.phys.nagoya-u.ac.jp}

\and

\author{MASAAKI TAKAHASHI}
\affil{Department of Physics and Astronomy, Aichi University of 
       Education, Kariya, Aichi 448-8542, Japan}
\email{takahasi@phyas.aichi-edu.ac.jp}

\begin{abstract}
 We present an analytical model for describing highly relativistic
 acceleration of magnetically driven jets, within the framework of 
 ideal MHD for cold, stationary and axisymmetric  outflows. Our novel
 procedure is to treat the wind equation as an algebraic relation
 between the relativistic Alfv\'{e}n Mach-number and  the poloidal
 electric to toroidal magnetic field amplitudes ratio $\xi$. 
 This allows us to obtain easily the wind solutions for 
 trans-fast-magnetosonic flows, together with the required range of 
 $\xi$.  Then, to determine  the spatial variation of $\xi$, we solve 
 approximately the Grad-Shafranov equation applied to a jet flow ejected 
 with a very large total specific energy $E$ and confined within a very
 small opening angle. Our trans-fast-magnetosonic model provides a
 closed-form expression for the transition from a magnetically dominated
 flow to a kinetic-energy dominated one, which occurs in the 
 sub-asymptotic region far beyond the light cylinder of the radius
 $R_{\rm L}$. Importantly, we find that the equipartition between magnetic
 and kinetic energies is realized at a cylindrical radius of order of
 $R_{\rm L}E/c^{2}$, and confirm that the further conversion of magnetic to
 kinetic energy proceeds logarithmically with distance in the asymptotic 
 region. Finally,  we discuss briefly the astrophysical implications of 
 our model for jets originating from active galactic nuclei. 
\end{abstract}

\keywords{MHD---relativity---galaxies: jets}

\section{INTRODUCTION}

\subsection{Collimated Relativistic Jets}

 Highly relativistic jet motion is one of the most interesting phenomena
 observed in active galactic nuclei (AGN), microquasars  and possibly
 $\gamma$-ray bursts (GRB), and many works have been devoted to numerical
 and analytical investigations of the mechanisms for producing,
 collimating and accelerating matter to relativistic speeds \citep{stt01,
 gc02}. At present the most promising approach to the jet phenomena seems
 to consider magnetically driven outflows within the framework of
 relativistic magnetohydrodynamics (MHD).

 Based on the MHD scenario, the jet ejection is expected to be realized
 under the magnetic energy-dominated state (i.e., the Poynting jet) in
 the vicinity of a central source. This initial Poynting flux will be
 originated by electromagnetic extraction of rotational energy from a
 spinning black hole or/and an accretion disk, as was first discovered
 by \citet{bz77} and numerically studied by the time-dependent MHD
 simulations in Kerr geometry (see Komissarov 2001 for the magnetically
 dominated regime; Koide et al. 2002 for the full MHD regime). 
 Then, a significant fraction of the huge energy in the outflow will be
 converted from the Poynting flux into the fluid kinetic energy of bulk
 motion. Such an efficient energy conversion may be able to occur
 unsteadily in the inner magnetosphere close to the black hole as 
 a result of the MHD interaction with infalling matter. \citet{ko00}
 discussed this problem and found a magnetically driven jet inside a gas
 pressure-driven jet in the counter-rotating black hole case against the
 disk rotation. However, the poloidal velocity of the jet is only
 sub-relativistic.

 Otherwise, the steady magneto-centrifugal acceleration in the
 propagation over a large enough distance should become important.  
 It is well-known that the ideal stationary axisymmetric MHD equations
 reduce to a set of two equations describing the local force-balance
 along the field and across the field, and called the poloidal wind
 equation and the Grad-Shafranov (trans-field) equation (for reference,
 see, e.g., Okamoto 1992; Beskin 1997). Self-similar solutions of
 magnetosphere, where the magnetic field lines are anchored to a thin
 accretion disk, were discussed for jet collimation and acceleration
 \citep{bp82,lcb92,co94}. Unfortunately, the outflows given by the 
 self-similar solutions \citep{co94} reach some maximum radius and
 recollimate afterwards. The self-similar scaling will not be valid in
 the asymptotic region. 
 Using the stationary axisymmetric model, the asymptotic flow structure
 has been found to vary logarithmically with radius, and the kinetic
 energy-dominated solutions describing collimating jet magnetospheres
 have been presented \citep{lcb92, ei93, bl94, to94}, while the
 decollimation of magnetic field lines has been claimed to become
 significant in the equatorial (current-closure) region \citep{bo00}. 
 Note that the asymptotic analysis fails to check whether the obtained 
 structure is due to trans-magnetosonic flows satisfying the critical 
 conditions. To assure the validity of the asymptotic structure, we 
 should consider the connection of the plasma source to the asymptotic 
 region (and to the black hole). However, the full MHD description of the 
 global magnetospheric structure in general relativity is a task still
 too complex even for the stationary axisymmetric system. For example,
 the available general relativistic models to give both closed and open
 field lines connecting a disk with a black hole and a distant region
 are limited to magnetic-energy dominated cases, such as a disk-current
 field \citep{tt01} and a force-free field \citep{fe97}.

 The jet-type outflows given by the one-dimensional or two-dimensional 
 numerical non-self-similar solutions, on the other hand, were also 
 constructed \citep{ac93a,ac93b,li93,fc96,fm01}. By solving the 
 Grad-Shafranov and poloidal equations self-consistently, disk-jet 
 connected flow solutions cylindrically confined by some external 
 pressure were obtained. To compare the MHD predictions with jet 
 observations in more detail ( e.g., the terminal velocity and width of
 the jets, the size of the active disk region in a central engine, which
 would correspond to jet's foot point, and the gravitational radius ),
 the efficiency of relativistic acceleration should be discussed in 
 more general asymptotic structures of outflows (depending on the 
 boundary conditions; e.g., the shape of the innermost flux surface, or 
 an external pressure distribution at the outermost flux surface, etc), 
 and the propagation distance necessary for a rough equipartition
 between magnetic and kinetic energies should be clarified. In this
 paper we would like to focus on such problems of steady acceleration at
 large distances from the central source, by developing a new analytical
 method without any self-similar assumption. Most of the previous works
 discussed the field structure under the force-free assumption, in which   
 the contributions by the plasma inertia are ignored. However, our new
 treatment allows us to study more easily trans-fast magnetosonic outflow 
 solutions and to include the plasma inertia effects on the magnetic 
 field structure and the plasma acceleration process. 
 Hence, by neglecting the effect of gravity and gas pressure, we can
 have some more self-consistent MHD solutions, showing the evolution of
 the bulk Lorentz factor $\gamma$ of outgoing flows to $\gamma \sim 10$
 in accordance with AGN jets.

\subsection{New Analytic Procedure for Trans-Magnetosonic Jets}

 In this paper, for the ideal stationary axisymmetric MHD outflows, 
 we follow the relativistic formulation given by \citet{cam86,cam87}. 
 To avoid a very troublesome task to solve the Grad-Shafranov equation,
 one may discuss the fluid motion by assuming a magnetic field
 configuration and solving only the poloidal wind equation \citep{ts98,
 fg01}. Though this procedure does not always assure the consistency
 with the Grad-Shafranov equation, the analysis of the poloidal wind
 equation is useful for studying the critical conditions for MHD flows. 
 For a cold plasma considered here, the key problem is to derive the
 trans-fast-magnetosonic solutions from the poloidal wind equation. If
 the poloidal wind equation is algebraically solved under a fixed field
 configuration in a conventional manner, however, it allows the
 existence of many unphysical solutions describing a flow which does not
 smoothly pass through the fast-magnetosonic critical point, unless the
 slightly complicated critical conditions are required for the four
 integrals of motion which are constant along a filed line. Usually this
 prevents us from understanding clearly the efficiency of acceleration
 along a field line in simple analytic computations.

 One of our purposes in this paper is to propose a new analytic
 procedure useful for discussing general properties of the trans-MHD
 flows within the poloidal wind equation and  without assuming a field
 configuration. We treat the poloidal wind equation as an algebraic
 relation between the relativistic Alfv\'{e}n Mach-number $M$ and the
 poloidal electric to toroidal magnetic field amplitudes ratio $\xi$
 (instead of the poloidal field amplitude). Though it is not difficult
 to contain the effect of gravity, our analysis given here is limited to
 special relativistic flows for showing explicitly the conversion of
 magnetic to kinetic energy at large distances. Then, it is easily found
 that only if $\xi$ is supposed to be a function smooth along a field
 line at the critical points, the solution (depending on $\xi$) for the
 poloidal wind equation becomes trans-critical. We would like to
 emphasize that no special critical conditions are needed for the four
 integrals of motion.

 The spatial variation of the ratio $\xi$ newly introduced into the
 poloidal wind equation must be determined by the Grad-Shafranov 
 equation. In the case of AGN (or possibly GRB) jets a high degree of
 collimation as well as a ultrarelativistic bulk motion is
 observationally indicated. Hence, we consider jet flows with a very
 small opening angle and a very large asymptotic Lorentz factor to solve
 approximately the Grad-Shafranov equation, under the condition of a
 cold plasma and no gravity. Hereafter we use the cylindrical
 coordinates $R$ and $Z$ with $c=1$ unit and denote the flux function by
 $\Psi(R,Z)$. For efficient acceleration of bulk motion a huge magnetic
 energy should be stored in the outflow near  the central source. Then,
 we assume the total specific energy $E(\Psi)$ to be much larger than 
 unity.  (In this paper, we use the term ``specific'' as a quantity per 
 the rest-mass energy of one plasma particle.)  The specific energy
 $E(\Psi)$ is conserved along a field line $\Psi=$ constant.
 The angular velocity of a field line $\Omega_{F}(\Psi)$ is also an
 integral of motion, and the light cylinder radius is given by
 $R_{\rm L}=1/\Omega_{F}$. By virtue of the very large value of $E$, 
 we  can define the intermediate region $R_{\rm L}\ll R\leq R_{\rm L}E$
 between the inner region $R\leq R_{\rm L}$ and the asymptotic region
 $R\gg R_{\rm L}E$.  
 We derive a class of general solutions corresponding to an arbitrary
 (i.e., cylindrical, paraboloidal or radial) boundary configuration for
 the outermost flux surface of jet flows with the opening angle smaller
 than $1/E$, by solving the approximated Grad-Shafranov equation, which
 remains valid in the spatial range from the intermediate region to the
 asymptotic region. The total specific energy $E$ may be decomposed into
 the magnetic part $E_{m}$ and the plasma kinetic part $E_{k}$. 
 The importance of the energy conversion from the magnetic part to 
 the kinetic part in a relativistic wind was originally pointed out by 
 \cite{mc69}, and he showed that for a radial magnetic field the energy 
 conversion from Poynting flux to particle energy flux (or the plasma 
 acceleration) is extremely inefficient. For a flux tube diverging 
 even slightly faster than radially, the significant energy conversion 
 occur beyond the fast magnetosonic point; at the fast magnetosonic
 point the total energy flux is still dominated by the Poynting flux 
 (e.g., \cite{bl94,ts98}). 
 The main purpose of this paper is to give the evolution of the ratio
 $E_{k}/E_{m}$ in a simple closed-form expression, and our new result 
 is to show the conversion of $E_{m}$ to $E_{k}$ in the intermediate
 region. The fast-magnetosonic critical point is found to exist in the
 intermediate region, where the flow is still magnetic-dominated
 ($E_{m}\gg E_{k}$) in accordance with the previous works. Importantly, 
 we can conclude that the rough
 equipartition $E_{m}\sim E_{k}$ is realized at a radius 
 $R\sim R_{\rm L}E$. The further energy conversion goes on
 logarithmically with the increase of $R$ toward the kinetic-energy
 dominated asymptotic state, as was previously pointed out \citep{ei93,
 bl94, to94}. The complete energy conversion becomes possible, if the
 outermost flux surface of jet flows extends to an infinite radius.

 This paper is organized as follows: In \S2, we analyze the poloidal
 wind equation by introducing the poloidal electric to toroidal magnetic
 field amplitudes ratio $\xi$. The general properties of the
 trans-fast-magnetosonic solutions are studied, in particular, for
 outflows which are initially ejected under a magnetic-energy dominated
 state and a very large total specific energy $E$. In \S3, we give the
 approximated form of the Grad-Shafranov equation for jet flows
 propagating beyond the light cylinder surface with a very small opening
 angle. In \S4, we derive a class of general solutions for the
 Grad-Shafranov equation to determine the Alfv\'{e}n Mach number $M$ and
 the flux function $\Psi$ in the intermediate and asymptotic
 regions. Finally we summarize our results of relativistic acceleration
 and discuss the implications for AGN jets in \S5.

\section{TRANS-CRITICAL MHD OUTFLOWS}

\subsection{Non-singular MHD flow equation}

 First let us give a brief review (see, e.g., Camenzind 1986) of ideal
 stationary axisymmetric MHD outflows for a cold plasma in Minkowski
 spacetime background with the cylindrical line element written by $c=1$
 unit as follows, 
\begin{equation}
 ds^{2}=dt^{2}-dR^{2}-dZ^{2}-R^{2}d\phi^{2} \ . \label{metric}
\end{equation}
 As was mentioned in \S1, we have the four integrals of motion which are
 constant along a field line given by $\Psi(R,Z)=$ constant. The total
 specific energy $E(\Psi)$ and angular momentum $L(\Psi)$ 
 are decomposed into the magnetic and kinetic parts as follows,  
\begin{equation}
 E=E_{m}+E_{k} \ , \ \ 
 E_{m}=-\frac{R\Omega_{F}B_{\phi}}{4\pi k} \ , \ \ 
 E_{k}=\gamma \ ,  \label{menergy}
\end{equation}
\begin{equation}
 L=L_{m}+L_{k} \ , \ \ 
 L_{m}=-\frac{RB_{\phi}}{4\pi k} \ , \ \ 
 L_{k}=\gamma R^{2}\Omega \ , \label{kenergy}
\end{equation}
 where the angular velocity of a field line $\Omega_{F}(\Psi)$ and the
 rest-mass energy loading rate per unit magnetic flux $k(\Psi)$ 
 are also the integrals of motion. (These four integrals of motion for
 outflows are assumed to be positive in the following.) We call
 $B_{\phi}$ the toroidal field, and the poloidal field strength is
 denoted by $B_{p}$. If the flux function $\Psi(R,Z)$ is determined, 
 the poloidal field strength is given by  
\begin{equation}
 B_{p}^{2}=(B^{R})^{2}+(B^{Z})^{2} \ , \ \ 
 B^{R}=-\frac{1}{R}\frac{\partial\Psi}{\partial Z} \ , \ \ 
 B^{Z}=\frac{1}{R}\frac{\partial\Psi}{\partial R} \ ,
\end{equation}
 and the toroidal field can be expressed in terms of the poloidal
 quantities. 
 Further, the Lorentz factor $\gamma$ equal to $E_{k}$ corresponds to
 bulk plasma motion involving both poloidal and rotational components. 
 The angular velocity of plasma is denoted by $\Omega$ in $L_{k}$. 
 The poloidal 4-velocity $u_{p}=\gamma v_{p}$ is used instead of the 
 3-velocity $v_{p}$. Then, the integral of motion $k$ is given by 
 $ k=\rho u_{p}/B_{p} $, where $\rho$ is the proper mass density of
 plasma.

 Now we can define the relativistic Alfv\'{e}n Mach number $M$ by the
 equation 
\begin{equation}
 M^{2} \equiv 4\pi \rho u_{p}^{2}/B_{p}^{2}=4\pi k u_{p}/B_{p} \ , 
                                                             \label{mach}
\end{equation}
 which is used to give the toroidal field as
\begin{equation}
 B_{\phi}=\frac{4\pi k}{M^{2}+x^{2}-1}
          \left(\frac{L\Omega_{F}}{x}-xE\right) \ ,  \label{toroidal}
\end{equation}
 and $x \equiv R\Omega_{F}=R/R_{\rm L}$. Because $\gamma$ and $R\Omega$  
 can be also represented by the four integrals of motion and $M^{2}$, 
 the condition $\gamma^{2}(1-R^{2}\Omega^{2})-u_{p}^{2}=1$ for
 normalization of 4-velocity leads to the poloidal wind equation   
\begin{equation}
 (1+u_{p}^{2})(1-x^{2}-M^{2})^{2}
 = e^{2}(1-x^{2}-2M^{2})
   +\left(E^{2}-\frac{L^{2}\Omega_{F}^{2}}{x^{2}}\right)M^{4} \ , 
 \label{wind}
\end{equation}
 where $e=E-L\Omega_{F}=\gamma(1-R^{2}\Omega\Omega_{F})$ is denoted 
 by the plasma parts of the conserved energy and angular momentum.  
 The angular velocity of plasma $\Omega$ is given by
\begin{equation}
 R\Omega = \frac{ (\Omega_F L/E) M^2 - x^2 (1-\Omega_F L/E) }
            { x (M^2 - 1 + \Omega_F L/E ) } \ .
\end{equation}

 Let us rewrite equation (\ref{mach}) to give the poloidal 4-velocity
 $u_{p}$ in equation (\ref{wind}) as follows, 
\begin{equation}
 u_{p}=B_{p}M^{2}/4\pi k \ . \label{vel}
\end{equation}
 Then, by assuming a specific flux function $\Psi(R,Z)$ (giving the 
 field strength $B_{p}$), one may solve the algebraic equation 
 (\ref{wind}) for $M^{2}$ to determine the evolution along a fixed field
 line. In general, however, such solutions giving $M^{2}$ as a function
 of $x$ and $\Psi$ would not be trans-fast-magnetosonic: If $u_{p}$
 becomes equal to the relativistic fast-magnetosonic wave speed at a
 point on a field line of $\Psi=$ constant, the partial derivative
 $\partial M^{2}/\partial x$ diverges there. (Otherwise, the flow
 remains sub-fast-magnetosonic or super-fast-magnetosonic everywhere on
 a field line.) It becomes necessary to find a special class of
 solutions satisfying the critical condition to keep a finite
 acceleration of $u_{p}$ at the fast-magnetosonic point (see,
 e.g., Takahashi \& Shibata 1998). In this paper we do not adhere to
 such an analysis of the critical condition relating the four integrals
 of motion with $B_{p}$ at the fast-magnetosonic point. We rather turn
 our attention to a different useful information concerning plasma
 acceleration of trans-fast-magnetosonic flows, which is derived from
 the poloidal wind equation without assuming any details of a flux
 function $\Psi(R,Z)$.

 For this purpose we introduce the poloidal electric to toroidal
 magnetic field amplitudes ratio $\xi$ as follows, 
\begin{equation}
 \xi\equiv \frac{E_{p}}{|B_{\phi}|}=x\frac{B_{p}}{|B_{\phi}|} \ , 
 \label{ratio}
\end{equation}
 where we have used the relation $E_{p}=xB_{p}$ between the poloidal
 field amplitudes. If equation (\ref{ratio}) by the help of equations
 (\ref{toroidal}) and (\ref{vel}) is used, the poloidal velocity $u_{p}$ 
 (and the Lorentz factor $\gamma$) is given by $M^{2}$ and $\xi$
 (instead of $B_{p}$). Then, the poloidal wind equation (\ref{wind}) 
 can be reduced to the quadratic equation for $M^{2}$ 
\begin{equation}
 AM^{4}-2BM^{2}+C=0 \ , \label{quadratic}
\end{equation}
 where the coefficients $A$, $B$ and $C$ are given by
\begin{eqnarray}
 A &=& E^{2}-1-\xi^{2}\left(E-\frac{L\Omega_{F}}{x^{2}}\right)^{2}
        -\frac{L^{2}\Omega_{F}^{2}}{x^{2}} \ ,  \\
 B &=& x^{2}+e^{2}-1 \ ,  \\
 C &=& -(x^{2}-1)B \ . 
\end{eqnarray}
 If the ratio $\xi$ is given as a smooth function of $x$ under a
 constant $\Psi$, the evolution of $M^{2}$ along a field line can be
 derived from equation (\ref{quadratic}). Then, we can determine the
 toroidal field $B_{\phi}$ from equation (\ref{toroidal}) and also the
 poloidal field $B_{p}$ from the relation $B_{p}=\xi |B_{\phi}|/x$. As
 will be briefly explained later, this field amplitude $B_{p}$ obtained
 through this procedure automatically satisfies the critical condition
 at the fast-magnetosonic point. This means that the quadratic equation
 (\ref{quadratic}) is not singular even at the fast-magnetosonic
 point. Hence, our necessary task in this section is just to solve
 explicitly equation (\ref{quadratic}) and to discuss the parametric
 range of $\xi$ for allowing the existence of highly relativistic
 trans-fast-magnetosonic solutions.

 After some algebraic manipulation, it is easy to show that equation
 (\ref{quadratic}) gives the solution of the form 
\begin{equation}
 M^{2}=\frac{x^{2}-1}{[E-(L\Omega_{F}/x^{2})]\sqrt{f/B}-1} \ ,  \label{sol}
\end{equation}
 where the function $f$ depending on $\xi$ and $x$ is defined by
 $ f \equiv (1-\xi^{2})x^{2}+\xi^{2} $.
 It is clear that the function $B$ involved in equation (\ref{sol})
 cannot be negative, because $e$ is estimated to be 
\begin{equation}
 e^{2}\geq\frac{(1-R^{2}\Omega\Omega_{F})^{2}}{1-R^{2}\Omega^{2}} \ .
\end{equation}
 Then, we must require the condition $f\geq 0$, under which we obtain
 the allowed range of $\xi$ for outflows propagating at a radius $x>1$
 beyond the light cylinder as follows, 
\begin{equation}
 \xi^{2}\leq\frac{x^{2}}{x^{2}-1} \ .
\end{equation}
 A more stringent constraint is obtained if we consider $M^{2}$ at the
 asymptotic radius $x\rightarrow\infty$, where we have 
\begin{equation}
 \frac{M^{2}}{x^{2}}\rightarrow\frac{1}{E\sqrt{1-\xi^{2}}-1} \ . 
\end{equation}
 This leads to the result that if the inequality
 $ \xi^{2}>\xi^{2}_{\rm c}\equiv 1-(1/E^{2}) $    
 always holds on a field line, $M^{2}$ becomes infinitely large at the
 radius $x=x_{\rm c}$ given by 
\begin{equation}
 x_{\rm c} = \frac{L\Omega_{F}}{1+E^{2}(\xi^{2}-1)}
 \left[\xi^{2}E-\frac{L\Omega_{F}}{2}
 +\sqrt{\xi^{2}(Ee-1)+\frac{L^{2}\Omega_{F}^{2}}{4}}\right]  \ . 
\end{equation}
 To avoid the divergence of $M^{2}$ at a finite radius, the field lines
 with $\xi>\xi_{\rm c}$ should become a closed loop or be asymptotically
 cylindrical within the radius $x=x_{\rm c}$. 
 Note that outside the light surface the closed loop configuration 
 should be forbidden because of the inertia effect of plasma. So the 
 flow streaming along the closed loop solution should transit to 
 another open field solution by making a MHD shock. Then, we can 
 expect that the asymptotically cylindrical configuration is a more 
 plausible one. (For the closed loop aligned flows without the shock 
 formation, the ideal MHD approximation may be broken at least near 
 the equatorial plane, because non-ideal MHD flows can stream across 
 the magnetic field lines.)

 The Alfv\'{e}n point on a field line is known to be present at 
 $M^{2}=M_{\rm AW}^2\equiv 1-x^{2}$, from which under the condition 
 $E>L\Omega_{F}$ the solution (\ref{sol}) gives the position  
 $ x^{2} = x_{\rm A}^{2} \equiv L\Omega_{F}/E $ 
 and the Alfv\'{e}n Mach number
 $ M^{2} = M_{\rm A}^{2} \equiv M_{\rm AW}^2(x_{\rm A}) = e/E $.
 In general, the differential form of the poloidal equation contains a 
 singular term (see, e.g., Takahashi \& Shibata 1998). However, with 
 the above condition, apparently the sub-Alfv\'{e}nic outflows given by
 the wind solution (\ref{sol}) can smoothly pass through the Alfv\'{e}n
 point and propagate to  the super-Alfv\'{e}nic region $x>x_{\rm A}$.  
 After passing through the light cylinder $x=x_{\rm L}=1$, which is 
 not a singular point for physical flows, the super-Alfv\'{e}nic
 outflows would reach distant regions. Of course, $M^{2}$ remains 
 finite also at the light cylinder, where we have  
\begin{equation}
 M^{2}_{\rm L} \equiv M^2(x_{\rm L}) = \frac{M_{A}^{2}}{1-w} \ , \ \ 
 w \equiv \frac{1+(\xi^{2}+1)e^{2}}{2eE} \ , 
\end{equation}
 The Alfv\'{en} Mach number at the light cylinder becomes larger 
 if compared with the value $M_{A}$ at the Alfv\'{e}n point.

 Now let us give the field amplitude $B_{p}$, using $M^{2}$ obtained 
 by equation (\ref{sol}). From the calculation of $B_{\phi}$ and the
 definition of $\xi$ we find the simple relation 
\begin{equation}
 B_{p}=\frac{4\pi k\xi}{M^{2}}\sqrt{\frac{B}{f}} \ .   \label{bp}
\end{equation}
 This allows us to estimate the field amplitude $B_{p}$ at the
 fast-magnetosonic point $x=x_{\rm F}$ with 
 $M^{2} = M_{\rm F}^2 \equiv M_{\rm FW}^2(x_{\rm F})$, where we have 
\begin{equation}
 M_{\rm FW}^2 \equiv \frac{f}{\xi^{2}} \              \label{fast}
\end{equation}
 for the Alfv\'{e}n Mach number related to the fast magnetosonic wave
 speed.  From equation (\ref{sol}) it is easy to find that the relation
 $M^2=M^2_{\rm F}$ is satisfied at the point determined by the equation 
\begin{equation}
 \left( E-\frac{L\Omega_{F}}{x^{2}} \right)^{2}f^{3} = x^{4}B \ . 
                                                     \label{position}
\end{equation}
 We must remark that the light cylinder radius $x=1$ satisfying equation
 (\ref{position}) is not the fast-magnetosonic point. In spite of the
 existence of such a redundant solution, we use this concise form
 (\ref{position}) for mathematical simplicity. In fact, equation
 (\ref{position}) is helpful for calculating the partial differentiation
 of equation (\ref{bp}) with respect to $x$ along a fixed field line at
 the fast-magnetosonic point, and we can straightforwardly check that
 the terms involving $\partial\xi/\partial x$ automatically cancel out
 in the derivative $\partial B_{p}/\partial x$ at this critical point,
 and the critical condition for the value of the partial derivative of
 $B_{p}$ holds, irrespective of any choice of $\xi$. (If the derivative
 $\partial\xi/\partial x$ diverges at the fast-magnetosonic point, the
 critical condition for $\partial B_{p}/\partial x$ is not satisfied. 
 This means that a smooth change of $\xi$, namely $B_\phi$, at the
 critical point is the essential requirement for a trans-magnetosonic 
 flow.) Thus, we can obtain a trans-fast magnetosonic flow solution
 without the regularity condition at the fast magnetosonic point. The
 behavior of the acceleration in the outgoing flow is determined by
 $\xi=\xi(x)$, and the asymptotic feature depends on the value of $\xi$;
 that is, $\xi>\xi_c$ or $\xi<\xi_c$.   
 The typical solutions for outgoing trans-fast magnetosonic flows 
 are demonstrated in Figures~\ref{fig:outflow}a and~\ref{fig:outflow}b, 
 for both cases of $\xi<\xi_c$ and $\xi>\xi_c$.   
 The outflows can start from the plasma source ($x\ll 1$), and after 
 passing through the Alfv\'{e}n point (A) and the fast magnetosonic 
 point (F) they reaches distant regions. In the case of $\xi<\xi_c$, 
 the flow reaches $x\gg 1$ region with a finite Mach number. This is 
 the standard picture discussed by lots of previous wind models. 
 On the other hand, in the case of $\xi>\xi_c$ the flow confines within 
 $x<x_c$. We can expect that asymptotically the magnetic field line 
 becomes cylindrical and the flow streams toward $Z$-direction. In the
 distant $Z$-region ($Z\Omega_F\gg 1$), the flow becomes to have a very 
 high Alfv\'{e}n Mach number.


\subsection{Acceleration and Energy conversion}

 Though we have constructed the basic formulae for discussing the
 evolution of $M^{2}$, in this paper we are particularly interested in
 highly relativistic acceleration of bulk motion through a conversion of
 magnetic to kinetic energy. In a sub-Alfv\'{e}nic region the outflows
 are expected to be injected under a magnetic-energy dominated state
 ($E\simeq L\Omega_{F}$) with a very large value of $E$. This means that
 we can analyze the evolution of $M^{2}$ in more details, using the
 approximations 
\begin{equation}
 \frac{e}{E}=\frac{E-L\Omega_{F}}{E}\ll 1 \ , \ \ E\gg 1 \ .
\end{equation}
 For example, equation (\ref{position}) claims that the value of
 $x^{2}_{\rm F}$ at the fast-magnetosonic point becomes very large and 
 is approximately given by $E^{2/3}f_{\rm F}$, in which the exact form 
 of $f=f(x)$ is necessary because $\xi_{\rm F} \equiv \xi (x_{\rm F})$ 
 may be very close to unity. Hence, we can write the position of the
 fast-magnetosonic point as follows,  
\begin{equation}
 x^{2}_{\rm F} 
 = \frac{\xi_{\rm F}^{2}E^{2/3}}{1+(\xi_{\rm F}^{2}-1)E^{2/3}} \ .
\end{equation} 
 We note that this critical point exists only if
 $\xi_{\rm F}^{2} > 1 - (1/E^{2/3})$, while the allowed range is 
 $1 < x_{\rm F} < \infty$, 
 depending crucially on the difference $\xi_{\rm F}^{2}-1$. On the other 
 hand, for the value of $M^{2}_{\rm F}$ at the fast-magnetosonic point 
 we obtain 
\begin{equation}
 \frac{M^{2}_{\rm F}}{x^{2}_{\rm F}}=\frac{1}{E^{2/3}}\ll 1 \ . \label{M/x}
\end{equation}

 To show the physical implication of equation (\ref{M/x}), we consider
 the specific magnetic energy $E_{m}$ at a radius $x\gg1$ beyond the
 light cylinder. From equations (\ref{menergy}) and (\ref{toroidal}) 
 we obtain approximately 
\begin{equation}
 E_{m}=\frac{E}{1+\mm^{2}} \ ,
\end{equation}
 where we have introduced the quantity $\mm^{2}$ defined by
 $ \mm^{2}\equiv M^{2}/x^{2} $. Then the ratio of the specific kinetic
 energy $E_{k}=E-E_{m}$ to $E_{m}$ is given by 
\begin{equation}
 \frac{E_{k}}{E_{m}}=\mm^{2} \ ,
\end{equation}
 which implies that $\mm^{2}$ is an indicator of the conversion of the 
 huge magnetic energy into the kinetic energy for outflows propagating 
 to a radius $x\gg 1$.  (The term $\mm^2$ corresponds to the inverse of
 the magnetization parameter $\sigma$ in some wind models; see, e.g,
 \cite{mc69} and \cite{cam86} for the radial wind, and \cite{bl94} and
 \cite{ts98} for non-radial winds.)  At the fast-magnetosonic point,
 however, we obtain from equation (\ref{M/x})  
\begin{equation}
 \left( \frac{E_{k}}{E_{m}} \right)_{\rm F} = \frac{1}{E^{2/3}}\ll 1 \ ,
\end{equation}
 which is a generic result  independent of a field configuration
 $\Psi(R,Z)$ of  highly relativistic outflows. It is clear that the
 energy conversion is still inefficient at the fast-magnetosonic point.

 Then, we consider plasma acceleration of trans-fast-magnetosonic
 outflows propagating at a radius $x\gg1$ beyond the critical point,
 where we have approximately 
\begin{equation}
 \mm^{2}=\frac{1}{\sqrt{(1-\xi^{2})E^{2}+(\xi^{2}E^{2}/x^{2})}-1} \ . 
 \label{largex}
\end{equation}
 As was previously mentioned, $M^{2}$ becomes infinitely large at
 $x=x_{\rm c}$ if $\xi>\xi_{\rm c}$. For highly relativistic outflows 
 this critical radius is approximately given by 
\begin{equation}
 x_{\rm c}^{2}=\frac{\xi^{2}E^{2}}{1+(\xi^{2}-1)E^{2}} \ .
\end{equation}
 A possible configuration of a field line given by the parameter $\xi$
 larger than $\xi_{\rm c}$ would be asymptotically cylindrical, because 
 it can be confined within a radius smaller than $x_{\rm c}$ even in the
 limit $Z\rightarrow\infty$. Unless $\xi$ is very close to unity, the
 value of $x_{\rm c}$ is of order of unity, and such a cylindrical jet
 becomes very narrow. In the limit $x\rightarrow x_{\rm c}$, the energy
 ratio $\mm^{2}$ is estimated to be 
\begin{equation}
 \mm^{2}=\frac{1}{1+(\xi^{2}-1)E^{2}}\times\frac{1}{1-(x/x_{\rm c})} \ ,
\end{equation}
 The narrow jet corresponding to $\xi_{\rm c}$ of order of unity can 
 be kinetic-energy dominated only if the asymptotic radius $x$ is very
 close to $x_{\rm c}$, i.e.,  $1-(x/x_{\rm c})=O(1/E^{2})$. This is 
 the fine-tuning problem required for the cylindrical field to realize
 highly relativistic acceleration of bulk motion. On the other hand for
 a cylindrical field line with the value of $\xi$ such that
 $|\xi^{2}-1|=O(1/E^{2})$ the maximum radius $x$ may become of order of
 $E^{2}$, and from equation (\ref{largex}) we find that at least a rough
 equipartition $E_{k}\sim E_{m}$ will be realized at the scale of
 $x\sim E$ without the fine-tuning of $x\rightarrow x_{\rm c}$. In
 particular, if $\xi\leq \xi_{\rm c}$, namely, $\xi^{2}\leq 1-(1/E^{2})$,
 the field line may be extending to an infinite radius
 $x\rightarrow\infty$, where $\mm^{2}$ becomes equal to
 $(E\sqrt{1-\xi^{2}}-1)^{-1}$. Such a field line configuration may be
 asymptotically paraboloidal or conical, and the efficient energy
 conversion into the state $E_{k}\geq E_{m}$ becomes possible at 
 $x\geq E$ for field lines with $1-\xi^{2}=O(1/E^{2})$. This is another 
 fine-tuning problem required for the poloidal electric and toroidal
 magnetic field amplitudes. The precise fine-tuning of
 $\xi^{2}=1-(1/E^{2})$ means that the magnetic energy $E_{m}$ can be
 completely transported into the kinetic energy $E_{k}$ as outflows
 propagate to an infinite radius.

 To claim that the kinetic energy can asymptotically become larger than
 the magnetic energy for injection of magnetic-energy dominated outflows
 with very large $E$, we must solve the fine-tuning problem such that
 $1-(x_{\rm c}/x)=O(1/E^{2})$ or $|\xi^{2}-1|=O(1/E^{2})$ as a result of 
 MHD interaction described by the Grad-Shafranov equation. In this paper 
 we consider a jet ejection with a very small opening angle such that
 $R/Z\leq 1/E$, and we study the spatial variation of $\xi$ to show the
 dynamical fine-tuning of $\xi$ in jet flows.

\section{THE APPROXIMATED GRAD-SHAFRANOV EQUATION}

 The asymptotic analysis of relativistic outflows has been developed in 
 previous works \citep{lcb92,ac93a,ac93b,ei93,bl94,to94}, and the
 logarithmic dependence of the asymptotic structure on $x$ has been
 pointed out. However, the transition from magnetic-energy dominated
 state into kinetic-energy dominated one is an important unsolved
 problem  which is beyond the usual scheme of the asymptotic analysis
 based on the naive approximation $x\gg 1$ in the poloidal wind and
 Grad-Shafranov equations. In our approach presented in the previous 
 section, such an approximation corresponds to  
\begin{equation}
 f\simeq (1-\xi^{2})x^{2}\gg 1 \ , 
\end{equation}
 for which we have
\begin{equation}
 \mm^{2}=\frac{1}{E\sqrt{1-\xi^{2}}-1} \
\end{equation}
 from equation (\ref{largex}). (This should be the case of $\xi\leq
 \xi_{\rm c}$.) The key point missed in this calculation is that the
 fine-tuning of $1-\xi^{2}=O(1/E^{2})$ to realize the energy
 equipartition ($\mm^{2}\sim 1$) may occur at a radius $x$ in the range
 $1\ll x\leq E$. Then, to discuss the energy conversion in the range
 $1\ll x\leq E$, which is called the ``intermediate'' range of $x$ in 
 this paper, we must analyze the evolution of $\mm^{2}$ without assuming
 $(1-\xi^{2})x^{2}$ to be very large. The existence of the intermediate
 range of $x$ is a main feature of highly relativistic outflows with a
 very large specific energy ($E\gg 1$). Recalling that we obtain
 $\mm^{2}\simeq e/E$ at the light cylinder surface $x=1$, we expect
 $\mm^{2}$ to increase from a sub-fast-magnetosonic value in the range
 $1/E\ll \mm^{2}\ll 1/E^{2/3}$ to a rough equipartition value of
 $\mm^{2}\sim 1$ as outflows propagate in the intermediate region. The
 fast-magnetosonic point at which we have the value of $\mm^{2}=1/E^{2/3}$
 can be involved in this region. The usual asymptotic analysis becomes
 valid only in the region $x\gg E$, where $\mm^{2}$ may increase
 logarithmically with $x$, and our purpose here is to give the more
 precise treatment valid both in the intermediate and asymptotic
 regions.

 The Grad-Shafranov equation can be written in the form
\begin{equation}
 \frac{M^{2}+x^{2}-1}{8\pi^{2}}\nabla\left(\frac{M^{2}+x^{2}-1}{R^{2}}
              \nabla\Psi\right)=s_{1}+s_{2} \ , \label{GS}
\end{equation}
 if no gravity is included (see, e.g., Tomimatsu 1994). The source terms
 $s_{1}$ and $s_{2}$ are given by 
\begin{equation}
 s_{1}=\frac{d}{d\Psi}(Ek)^{2}-\frac{1}{R^{2}}\frac{d}{d\Psi}(Lk)^{2}
      -\frac{1}{M^{2}}\frac{d}{d\Psi}(ek)^{2} \ ,
\end{equation}
 and
\begin{equation}
 s_{2}=-\frac{M^{2}+x^{2}-1}{M^{2}}\frac{d}{d\Psi}k^{2}
       -\frac{R^{2}k^{2}}{M^{4}}(M^{2}+x^{2}-1+e^{2})
        \frac{d}{d\Psi}\Omega_{F}^{2} \ . 
\end{equation}
 Here we regard $M^{2}$, $\partial\Psi/\partial R\equiv\Psi_{R}$ and
 $\partial\Psi/\partial Z\equiv\Psi_{Z}$ in equation ({\ref{GS}) as
 functions of the variables $R$ and $\Psi$, instead of $R$ and
 $Z$. Further, by virtue of the introduction of $\xi$, from equations
 (\ref{sol}) and (\ref{bp}) we find the relation 
\begin{equation}
 M^{2}+x^{2}-1 = 4\pi k\left(E-\frac{L\Omega_{F}}{x^2}\right)
                 \frac{\xi}{B_{p}} \ , \label{bp2}
\end{equation}
 which is substituted into the left-hand side of equation
 (\ref{GS}). Then, after some manipulation we obtain 
\begin{equation}
 \frac{1}{\Psi_{R}}\frac{\partial}{\partial R}
  \left[\frac{\xi^{2}k^{2}}{1+q^{2}}\left(E-\frac{L\Omega_{F}}{x^2}\right)^{2}
  \right] = \tilde{s}_{1}+s_{2} \ , \label{mGS}
\end{equation}
 where $q \equiv -\Psi_{Z}/\Psi_{R}$ means the slope of a poloidal
 magnetic field line, and the modified source term $\tilde{s}_{1}$ is   
\begin{equation}
 \tilde{s}_{1} = -\frac{\partial}{\partial\Psi}
 \left[\left(E-\frac{L\Omega_{F}}{x^2}\right)^{2}k^{2}\xi^{2}\right] 
  + s_{1} \ .
\end{equation}

 Now we derive the approximated form of the Grad-Shafranov equation
 valid for highly relativistic outflows (with $E\gg e\geq1$) propagating
 in the intermediate and asymptotic regions, where we obtain 
\begin{equation}
 1 \gg    1-\xi^{2}+\frac{\xi^{2}}{x^{2}}
   \simeq \frac{1}{E^{2}}\left( 1+\frac{1}{\mm^{2}} \right)^{2} \ .
\end{equation} 
 as a result of the approximation $M^{2}+x^{2}-1\simeq (\mm^{2}+1)x^{2}$
 in equations (\ref{bp2}) and (\ref{bp}). It is interesting to note that
 the source terms are reduced to the compact form 
\begin{equation}
 \tilde{s}_{1}+s_{2}\simeq (1+\mm^{2})\Omega_{F}^{2}
 \frac{\partial}{\partial\Psi}
 \left(\frac{k^{2}}{\Omega_{F}^{2}\mm^{4}}\right) \ .
\end{equation}
 To see clearly the fine-tuning of $\xi$, we rewrite it as follows
\begin{equation}
  \xi^{2} \simeq 1-\frac{\beta}{E^{2}} \ ,
\end{equation}
 where we define 
\begin{equation}
 \beta \equiv   
 \left(1+\frac{1}{\mm^{2}}\right)^{2} -\frac{1}{\y^{2}} \ . \label{mbeta}
\end{equation}
 The renormalized variable $\y \equiv x/E$ is also useful to discuss 
 the evolution of $\mm^{2}$ in the regions considered here.

 The observations of AGN jets (and possibly GRB jets) show the existence
 of highly collimated outflows with very narrow opening angles, for
 which we can assume that $q^{2}\ll 1$ in equation
 (\ref{mGS}). Hereafter, we consider only the case of $q<1/E$ to give 
\begin{equation}
 R\Psi_{R}\simeq R^{2}B_{p} 
 \simeq\frac{4\pi kE}{\Omega_{F}^{2}(1+\mm^{2})} \ . \label{mpsi}
\end{equation}
 This allows us to eliminate $\Psi_{R}$ from the left-hand side of
 equation (\ref{mGS}), in which we also have
 $[E-(L\Omega_{F}/x)]^{2}\simeq E^{2}-(2/\y^{2})$. Then, we arrive at
 the final form of the reduced Grad-Shafranov equation  
\begin{equation}
 R\frac{\partial\beta}{\partial R} = 
 -\frac{4\pi E}{k} \frac{\partial}{\partial\Psi}
  \left(\frac{k^{2}}{\Omega_{F}^{2}\mm^{4}}\right)
 +\frac{4}{\y^{2}} \ . \label{apGS}
\end{equation}
 If equation (\ref{mbeta}) is substituted into equation (\ref{apGS}), 
 we can obtain the partial differential equation of first order for
 $\mm^{2}$ as follows,  
\begin{equation}
 \frac{R(\mm^2+1)}{\mm^6} \frac{\partial \mm^2}{\partial R} 
 + \frac{b}{\mm^6} \frac{\partial \mm^2}{\partial\Psi} 
 =  \frac{a}{\mm^4} - \frac{1}{\y^{2}}  \ ,        \label{basic}
\end{equation}
 where 
\begin{equation}
 a=\frac{4\pi E}{\Omega_{F}}\frac{\partial }{\partial\Psi}
 \left(\frac{k}{\Omega_{F}}\right) \ , 
\end{equation}
 and
\begin{equation}
 b=\frac{4\pi Ek}{\Omega_{F}^{2}} \ .
\end{equation}
 In the intermediate region ($\y\leq 1$) the role of the term $1/\y^{2}$
 in equation (\ref{basic}) becomes quite important, while it has been
 completely neglected in the asymptotic analysis corresponding to the
 case of $\y\gg 1$. If a solution $\mm^2(R, \Psi)$ is derived from 
 equation (\ref{basic}), it is straightforward to find the flux function
 $\Psi(R,Z)$ from equation (\ref{mpsi}). In the next section we will
 present the interesting example of a jet solution to discuss the energy
 conversion and the change of field configuration in the intermediate
 and asymptotic regions.

\section{FIELD STRUCTURE OF JET FLOWS}

 Equation (\ref{basic}) is valid only for jets with small opening angles
 propagating beyond the light cylinder surface. Considering the limit
 $\Psi\rightarrow 0$ for such outflows, we expect the jet ejection to
 occur near the polar axis $R\Omega_{F}\rightarrow 0$, where the
 toroidal filed $B_{\phi}$ should decrease in proportion to $R$. Noting  
 that the small flux function $\Psi$ is also proportional to $R^{2}$, 
 we estimate to be $E \sim E_m \sim \Psi/k$ in the injection region 
 near the polar axis. Then, to obtain a large specific energy $E$ per
 one particle, the rest-mass energy loading rate $k$ per unit magnetic
 flux should also become small in proportion to $\Psi$. This boundary
 condition near the polar axis motivates us to consider the case such
 that $E$ and $\Omega_{F}$ are independent of $\Psi$, while 
 $k=k_{0}(\Psi/\Psi_{0})$, where $k_0$ and $\Psi_0$ are constants and
 should be given by the boundary conditions at some foot point in a
 plasma source.  Then, we obtain  
\begin{equation}
 b=a\Psi \ , \ \ a=\frac{4\pi Ek_{0}}{\Omega_{F}^{2}\Psi_{0}}  \ ,
\end{equation}
 where $a$ is a dimensionless constant.

 Further, let us recall that $\mm^{2}\sim 1/E$ near the light cylinder
 surface, Then, we expect $\mm^{2}$ to be very small in proportion to
 $\y$ in the region $1/E\ll \y\ll 1$, from which equation (\ref{mpsi})
 leads to $R\Psi_{R}\simeq a\Psi$. This means that for $a=2$ we have
 $\Psi\propto R^{2}$, assuring the smooth matching to the inner solution
 valid near the polar region. (Of course, we cannot give the condition
 $a=2$ to be necessary, because our analysis is limited to the outer
 solution valid in the range $R\Omega_{F}\gg 1$.) Fortunately, for the
 model with $a=2$ and $b=2\Psi$, we can give the general solution of an
 analytical form for equation (\ref{basic}) as follows, 
\begin{equation}
 \frac{2\y^{2}(1+\mm^2)}{2\y^{2}+\mm^2} 
  = \ln\left( \frac{2\y^{2}}{\mm^2}+1 \right) + D_{1} \ , \label{sol1}
\end{equation}
 where $D_{1}$ is an arbitrary function of $D_{2}$ defined by
\begin{equation}
 D_{2} \equiv \frac{\Psi}{\Psi_{0}} 
       \left( \frac{1}{\mm^2} + \frac{1}{2\y^{2}} \right) \ .   \label{sol2}
\end{equation}
 Applying equation (\ref{mpsi}) to this general solution, we find that
 $D_{1}$ and $D_{2}$ are arbitrary functions of $Z$, which should be
 determined by the additional boundary conditions for jet flows. We can
 study an essential feature of plasma acceleration in jet propagation
 from this simple model.

 First let us discuss the results obtained from equation (\ref{sol1}). 
 Considering that $\mm^{2}\sim 1/E$ near the light cylinder
 $\y_{\rm L}=1/E$, we claim that $\y^{2}/\mm^{2}\sim \y \ll 1$ in the 
 range $1/E < \y \ll 1$, from which equation (\ref{sol1}) leads to  
\begin{equation}
 2\y^{2}(1-\y^{2}/\mm^4)\simeq D_{1}(Z) \ . \label{sol3}
\end{equation}
 for all $Z$. This equation means that the absolute value of $D_{1}(Z)$
 should be chosen to be at most of order of $1/E^{2}$ for all $Z$. In
 fact, if  $|D_{1}|\gg 1/E^{2}$ for some $Z$, we can easily see that
 equation (\ref{sol3}) breaks down in the range $1/E^{2}\ll \y^2 \ll
 |D_{1}|$. Hence, we can discuss the increase of $\mm^{2}\simeq
 E_{k}/E_{m}$ in the range $\y \gg 1/E$ under the choice of $D_{1}=0$,
 namely, according to the following equation 
\begin{equation}
 \frac{2\y^{2}(\mm^{2}+1)}{2\y^{2}+\mm^{2}} = 
 \ln\left(\frac{2\y^{2}}{\mm^{2}}+1\right) \ .  \label{model}
\end{equation}
 It is now clear that we have $\mm^{2}\simeq \y$ in the range 
 $1/E \ll \y \ll 1$, and the outflows can pass through the
 fast-magnetosonic point (i.e., $\mm^{2}=1/E^{2/3}$) located on the
 radius $R\Omega_{F}\simeq E^{1/3}$. If the outflows can arrive at the
 radius $\y \simeq 1.4$, the equipartition $E_{k}=E_{m}$ between
 magnetic and kinetic energies is realized. In the asymptotic region 
 $\y \gg 1$ we can confirm the logarithmic increase of $\mm^{2}$ given
 by $\mm^{2}\simeq\ln(2\y^{2}/\mm^{2})$ toward the complete conversion 
 of magnetic to kinetic energy.  
 The numerical solution of equation (\ref{model}) is shown in Figure 
 \ref{fig:gs-mm}a. The corresponding poloidal velocity $u_p(\y)$ and 
 Lorentz factor $\gamma(\y)$ are also shown in Figure \ref{fig:gs-mm}b, 
 where the Lorentz factor is given by $\gamma = E - u_p/(\xi \mm^2)$.  
 In this case, we obtain $\xi(\y)<1$ anywhere for the ratio of the 
 poloidal electric to toroidal electric magnetic field amplitude (see 
 Fig.\ref{fig:gs-mm}c).  
 Note that the Lorentz factor includes both the poloidal and toroidal 
 motion. In Fig.\ref{fig:gs-mm}b, the difference between $u_p$-value 
 and $\gamma$-value near the light cylinder is due to the dominated 
 toroidal motion of the plasma. (Just around the light cylinder 
 $\y \sim \y_{\rm L}$, the properties of the flow shown in Fig.
 \ref{fig:gs-mm} may be incorrect because of our approximations, but 
 we can expect correct features at least near and outside the fast 
 magnetosonic point, $\y \geq \y_{\rm F}$. )


 Next, we discuss the field configuration given by equation
 (\ref{sol2}). The jet flows may be confined by an external pressure 
 (see, e.g., Li 1993; Begelman \& Li 1994; Fendt 1997). If the shape
 $Z\Omega_{F}=H(\y)$ of the last flux surface $\Psi=\Psi_{0}$ is
 determined by the outer boundary condition, the function $D_{2}(Z)$ in
 equation (\ref{sol2}) is fixed. For example, let us consider the radial
 last flux surface at an angle $R/Z \equiv \theta_{0}$ with the pole
 axis direction. (Because the solution can be applied only to jets with
 small opening angles, we must require that $\theta_{0}$ is at most of
 order of $1/E$.) Using the function $\mm^2=\mm^2(\y)$ derived from
 equation (\ref{model}), we obtain  
\begin{equation}
 D_{2}(\theta_0 \z) = \frac{1}{\mm^2(\theta_0 \z)} 
              + \frac{1}{2(\theta_0 \z)^2} \     \label{eq:d_Z}
\end{equation}
 along a flux function $\Psi(R,Z)=$ constant, where $\z \equiv
 Z\Omega_F/E$. Note that for the value of $\mm^2(\theta_0 \z)$ in
 equation (\ref{eq:d_Z}) we use the function $\mm^2=\mm^2(\y)$, where   
 the variable $\y$ should be replaced to $\theta_0 \z$. Then, in the
 range $\theta_{0}Z\Omega_{F}/E \ll 1$ of $Z$, equations (\ref{sol2})
 and (\ref{eq:d_Z}) leads to the conical field configuration 
\begin{equation}
 \frac{Z}{R} \simeq \frac{1}{\theta_{0}}\sqrt{\frac{\Psi_{0}}{\Psi}} \ .
\end{equation}
 This shape of field lines changes as $Z$ increases, and in the
 asymptotic region we find the leading behavior such that 
\begin{equation}
 \frac{\theta_{0}Z\Omega_{F}}{E}
 \simeq \left(\frac{R\Omega_{F}}{E}\right)^{\Psi_{0}/\Psi} \ , \label{par}
\end{equation}
 which represents a paraboloidal collimation of the inner field lines
 for $\Psi<\Psi_{0}$. Figure \ref{fig:z-x} shows the shape of the
 magnetic field lines, which can be solved numerically by using
 equations (\ref{eq:d_Z}) and (\ref{sol2}) with the function
 $\mm^2=\mm^2(\y)$ ( or $\mm^2=\mm^2(\theta_0 \z)$ ) derived from
 equation (\ref{model}).


 From the solution (\ref{model}) we note that the ratio $\mm^{2}$ of
 kinetic to magnetic energy is completely determined by the propagation
 radius $\y=R\Omega_{F}/E$, irrespective of the field configuration given
 by equation (\ref{sol2}). If the last flux surface is asymptotically
 cylindrical with the outermost radius $\y_{0}$, giving the function
 $D_{2}(Z)$ approaching a constant value in the limit
 $Z\rightarrow\infty$, the maximum value of $\mm^{2}$ in the cylindrical
 jets is crucially dependent on $\y_{0}$. For narrower jets with
 $\y_{0}<1$ the magnetic energy is still dominant even in the asymptotic
 region.

 The outer flux surface corresponds to a larger radius $\y$ if compared
 with a fixed vertical distance $Z$. Thus, we can claim also that the
 energy conversion along outer flux surfaces of larger $\Psi$ becomes
 more efficient. This tendency of the efficient energy conversion is due
 to the condition for the rest-mass energy loading rate per unit
 magnetic flux such that $k\propto\Psi$. In fact, if we consider the
 case that $E$, $\Omega_{F}$ and $k$ are independent of $\Psi$, we find
 a different dependence of $\mm^2$ on $\y$ from the general solution for
 equation (\ref{basic}) with $a=0$ and a constant $b$, which is given by 
\begin{equation}
 \left( \frac{1}{\mm^2}+1 \right)^{2}+\frac{1}{\y^{2}}
  = F_{1}^{2} \ , \label{constb}
\end{equation}
 and
\begin{equation}
 \exp\left(\frac{2\Psi}{b}\right) 
  = \y^{2}F_{2}^{2} \left(\frac{\mm^2 F_{1}-\mm^2-1}
                   {\mm^2 F_{1}+\mm^2+1}\right)^{1/F_{1}} \ .
\end{equation}
 From equation (\ref{mpsi}) we can check that $F_{1}$ and $F_{2}$ are
 arbitrary function of $Z$ . If we assume the behavior of these
 functions in the limit $Z\rightarrow\infty$ such that $F_{1}\simeq
 1+(1/\ln Z)$ and $F_{2}$ remains finite, we can confirm the increase 
 of $\mm^{2}$ in logarithmic scale of $Z$ and the paraboloidal shape 
 of  field lines similar with equation (\ref{par}) in the asymptotic
 region. However, it is clear from equation (\ref{constb}) that 
 $\mm^{2}$ decreases as $\y$ increases under a fixed
 $Z$. The energy conversion becomes less efficient on outer flux
 surfaces. Further we note that the solution (\ref{constb}) fails to
 give a real value of $\mm^2$ in the range $\y \ll 1$ under a fixed $Z$. 
 Such a region should be covered by inner flux surfaces corresponding 
 to small $\Psi$, and the previous solution (\ref{model}) with
 $k\propto\Psi$ should be used there. One may consider a more realistic
 dependence of $k$ on $\Psi$. Then, a smooth change of $\mm^2$ from
 equation (\ref{model}) to equation (\ref{constb}) will be allowed as
 $\Psi$ increases in the range $0 < \Psi < \Psi_{0}$,

 It is sure that according to a choice of the integrals of motion as
 functions of $\Psi$ we can obtain various evolutionary models of energy
 conversion in jet flows, which may show more complicated behaviors
 different from the simplest solution (\ref{model}). However, we would
 like to emphasize the key result obtained here that a rough
 equipartition between magnetic and kinetic energies is realized at the
 radius $R \sim R_{\rm L}E$ far beyond the light cylinder radius
 $R_{\rm L}$,  and the subsequent logarithmic increase of kinetic energy
 goes on at larger radii $R \gg R_{\rm L}E$. This potentiality of MHD
 acceleration will be robust at least under the boundary condition for
 inner flux surfaces $\Psi \rightarrow 0$ such that the rest-mass energy 
 loading rate is given by $k \propto \Psi$ to keep the total specific 
 energy $E$ very large. If so, we can claim that there exists the
 critical scale given by $R = R_{\rm L}E$ for conversion of Poynting flux
 injected in jet flows, which is important for discussing a prompt
 emission of radiation owing to dissipation of kinetic energy of bulk
 motion.

\section{SUMMARY AND DISCUSSION}

 We have presented a parametric representation of Alfv\'{e}n Mach number
 $M$ by the ratio $\xi$ of poloidal electric to toroidal magnetic field
 strength, to avoid the troublesome analysis of the critical condition
 at the fast-magnetosonic point. Then, from the poloidal wind equation
 we have easily derived trans-fast-magnetosonic solutions including the
 parameter $\xi$ defined as a smooth function along a field line. If the
 parametric function $\xi$ is asymptotically larger than the critical
 value $\xi_{\rm c}$, the flux surface have been shown to be confined 
 within a finite radius $x=x_{\rm c}$. Otherwise, the dynamical
 fine-tuning of $\xi$ have been required for acceleration to highly
 relativistic bulk speeds. To determine the parametric function $\xi$,
 we have also given the approximated form of the Grad-Shafranov equation
 applied to jets ejected with a very large total specific energy $E$ and
 confined within a very small opening angle of order of $1/E$.

 Acceleration of outflows in generic models with the integrals of motion
 $E$, $\Omega_{F}$ and $k$ variously dependent on $\Psi$ has not been
 analyzed in this paper. The comparison of the two models given by
 equations (\ref{model}) and (\ref{constb}) suggests a variability of
 acceleration efficiency depending on various choices of the integrals
 of motion. Nevertheless, the former solution (\ref{model}) studied in
 the previous section will be interesting as a typical model revealing
 the high potentiality of MHD acceleration, which was also discussed by
 \citet{ok02} in relation to pulsar winds (see also \cite{mc69,bl94}). 
 By virtue of  the closed-form expression (\ref{model}) of the model, 
 we can clearly understand the following evolution of jet flows in the
 intermediate and asymptotic regions far beyond the light cylinder, if 
 the jet radius $R$ extends to an infinite distance:  
 (1) The magnetic-energy dominated and sub-fast-magnetosonic outflows
 pass through the light cylinder surface ($R=R_{\rm L}$) with the
 Alfv\'{e}n Mach number such that $M^{2}\sim1/E$. (2) Then, the energy
 ratio $E_{k}/E_{m}$ increases to $1/E^{2/3}$ at the fast-magnetosonic
 point corresponding to the radius $R\sim R_{\rm L}E^{1/3}$. In the
 intermediate region ($R_{\rm L} \ll R \leq R_{\rm L}E$) the outflows
 can smoothly become super-fast-magnetosonic. (3) Importantly, the model
 claims the realization of rough equipartition between kinetic and
 magnetic energies at the radius of order of $R_{\rm L}E$. (4) The
 further energy conversion toward a kinetic-energy dominated state in
 logarithmic scales of $R$ in the asymptotic region ($R \gg R_{\rm L}E$)
 is also confirmed. Figure \ref{fig:jet} is the summary of this jet
 solution.


 The full conversion of magnetic energy into kinetic one means that the 
 asymptotic Lorentz factor of bulk motion becomes equal to the total  
 specific energy $E$ of outflows injected near the central source. 
 Hence, for kinetic-energy dominated jets observed with a huge bulk 
 Lorentz factor $\gamma$, we can expect the rough equipartition 
 $E_{k}\sim E_{m}$ to occur at the jet radius 
 $R_{\rm jet}\equiv R_{\rm L}\gamma$   
 [ Note that the value of $\gamma$ ($\sim E_k$) is roughly same order
 with $E$ ].  Though we have considered ideal MHD flows in this paper, 
 the observed jet activity, such as a prompt emission of radiation 
 and a ultra relativistic acceleration of electrons, should be due to
 dissipation of the power of bulk motion, for example, through formation
 of shocks. Then, the interesting high-energy phenomena of jets will be
 observed, only after the kinetic energy of bulk motion begins to 
 dominate, namely, the jet radius extends to this critical radius
 $R_{\rm jet}$. 
 The shock formed in the energy equipartition region would be distinct 
 from shocks formed in the kinetically dominated asymptotic region by 
 observations. For the kinetically dominated flows, the compression
 ratio behind the shock is much higher, and flatter synchrotron spectra, 
 higher emissivities, etc. would be observed \citep{bl94}.

 In particular, for AGN jets with $\gamma\sim 10$ 
 (see, e.g., Ghisellini et al. 1993),  
 we can estimate the critical radius to be 
 $R_{\rm jet} \sim 10R_{\rm L}$. When magnetic fluxes for the jet
 connect to a rotating geometrically thin disk around a black hole, we
 can regard $\Omega_F$ as $\Omega_{\rm K}(r)$, where $\Omega_{\rm K}(r)$
 is the angular velocity for circular equatorial orbit in the Kerr
 metric, which corresponds to the Keplerian angular velocity in the
 Newtonian case. If the foot points of most magnetic fluxes distribute
 near the inner part of the disk $ r \sim r_{\rm ms}$, the angular
 velocities of the magnetic field lines are roughly 
 $\Omega_F(\Psi) \sim \Omega_{\rm K}(r_{\rm ms})$ 
 (see, e.g., Camenzind \& Krockenberger 1992).  
 Then, the light cylinder radius $R_{\rm L}=1/\Omega_F$ for field lines  
 threading a rotating black hole and the disk's inner part is possibly 
 equal to several Schwarzschild radii. Hence, our conjecture from the
 jet model obtained here is that the active region in AGN jets, far 
 from the central source, has the radius as large as a hundred of
 Schwarzschild radii, corresponding to $0.01$ pc for a 
 $10^{9} M_{\odot}$ black hole.  This result is consistent with
 observational results of blazars \citep{kataoka01} and of the M87 jet
 \citep{jbl99}. (For GRB jets expected to have higher Lorentz factors
 $\gamma>100$ of bulk motion, the size of the emission region will be
 larger than a thousand of Schwarzschild radii.)  The dependence of 
 the radius of the emission region on the bulk Lorentz factor is an 
 interesting problem to be checked by observations of AGN jets.

 Finally let us discuss the vertical distance $Z_{\rm jet}$ from the
 central source, which gives the critical radius $R_{\rm jet}$ on a
 field line. We have postulated jet flows confined within a small
 opening angle of order of $1/E$ and derived the solution valid only in
 the region $Z\geq ER$ from the Grad-Shafranov equation. We can obtain
 the field configuration according to equation (\ref{sol2}) under a
 suitable boundary condition on the outermost flux surface. As an
 example, in the previous section, we have considered the case with the
 radial outermost flux surface given by $Z=R/\theta_{0}$. If the opening
 angle $\theta_{0}$ is equal to $1/\gamma$, we can claim that the 
 kinetic energy of bulk motion begins to dominate at the vertical 
 distance $Z_{\rm jet}$ larger than  
 $\gamma R_{\rm jet} = \gamma^{2}R_{\rm L}$, far from the central
 source. (The value of $Z_{\rm jet}$ becomes larger on inner flux
 surfaces corresponding to smaller $\Psi$.) This will predict a
 sub-parsec distance to the active region in an AGN jet from a  
 $10^{9} M_{\odot}$ black hole. We expect that at this distant region
 ($Z>Z_{\rm jet}$) internal shocks form and make the active region in
 the jet. The internal shock scenario is generally thought for gamma-ray
 bursts, radio-loud quasars and blazars 
 (see, e.g., Mezaros \& Rees 1993; Spada et al. 2001; Ghisellini
 \& Celotti 2002). 
 In this scenario, it is assumed that the active region powered by
 collisions among different part of jet itself, moving at different bulk
 Lorentz factors. Although we consider stationary flows in the
 magnetosphere, we can also expect some kinds of plasma instability
 (e.g., the screw instability of plasma in black hole magnetosphere
 discussed by \cite{tmt01}) to make blobs of plasma in the jet.  If the
 cylindrical collimation of the outermost flux surface develops in the
 range $R < R_{\rm jet}$, however, the vertical scale $Z_{\rm jet}$
 should be much larger. Because we have not discussed physical
 mechanisms (possibly due to external matter) of the confinement of the
 magnetic flux, the boundary condition on the outermost flux surface to
 determine the field configuration remains arbitrary. This is a problem
 beyond the scope of the present paper, which is important for giving a
 more definite estimation of the vertical distance from the central
 source to the active region in AGN jets.

\acknowledgments

 This research was partly supported by Scientific Research Grant (C2)
 14540253 of Japan Society for Promotion of Sciences.


\begin{figure}
   \epsscale{0.75}
   \plotone{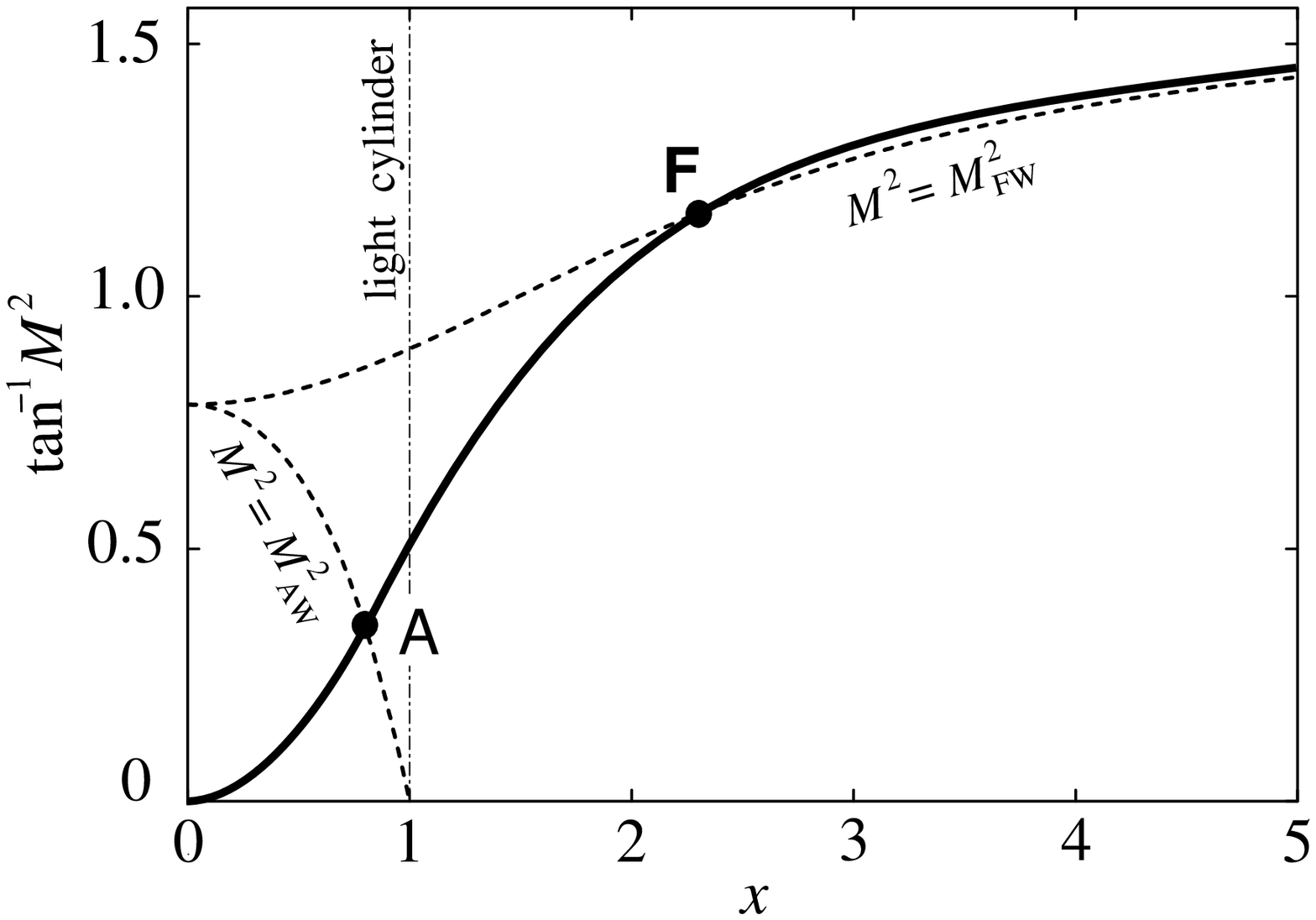} 
   \vspace{5mm}
   \plotone{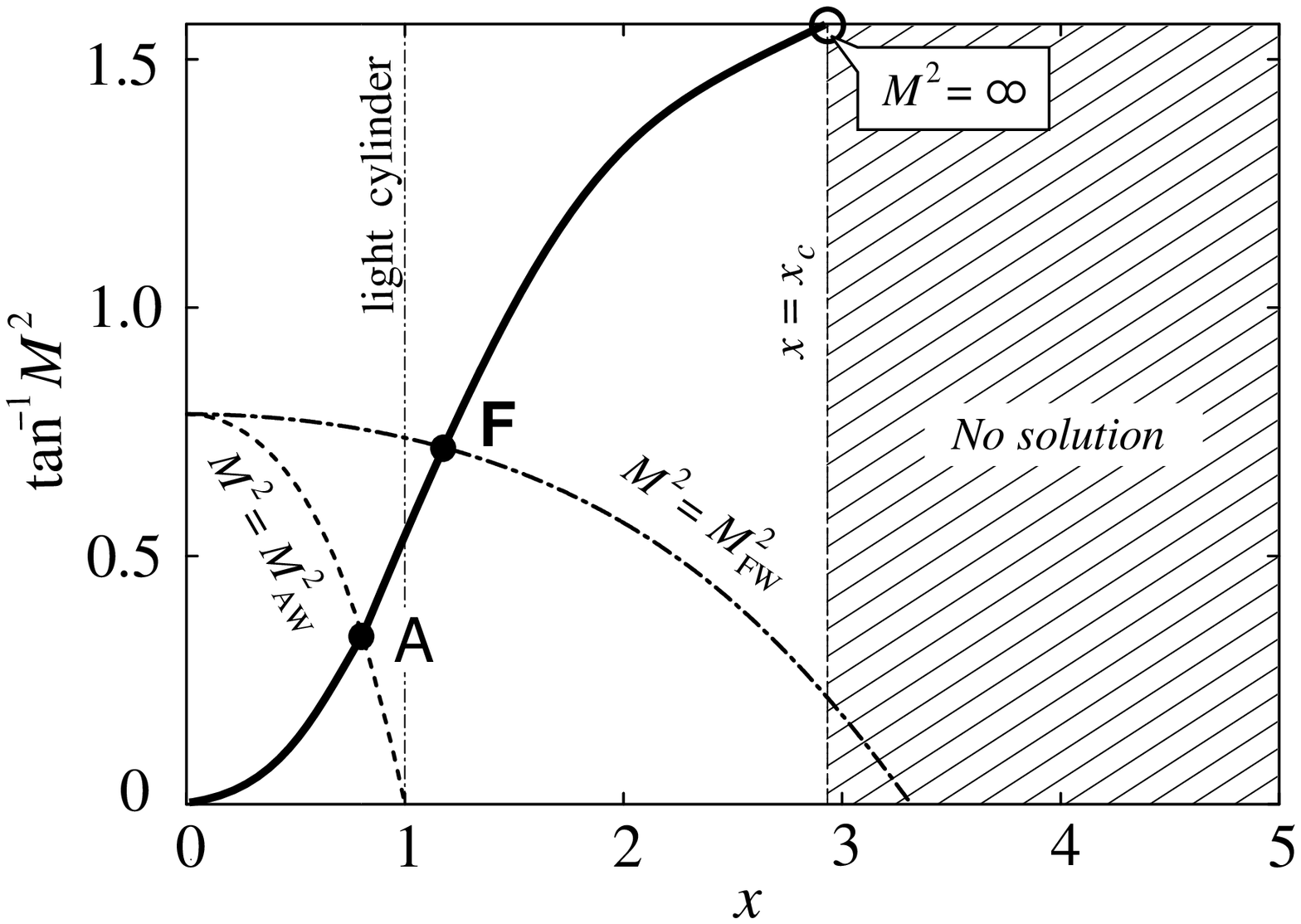} 
\caption{ 
 Examples of outgoing flow solutions (thick curves) given by equation (15) 
 with (a) $\xi=0.8$ and (b) $\xi=1.1$, where $E=10.0$ and $x_{\rm A}=0.8$. 
 The curves of the Mach numbers corresponding to the Alfv\'{e}n wave
 speed  and the fast-magnetosonic wave speed, namely, $M^2=M^2_{\rm AW}$
 and $M^2=M^2_{\rm FW}$, are also plotted, where $M^2_{\rm AW}=1-x^2$ 
 and $M^2_{\rm FW}=1-x^2+(x^2/\xi^2)$. The crossings of these curves
 with the flow solution labeled by ``A'' and ``F'' are the Alfv\'en and
 fast-magnetosonic points, respectively.  
        } 
\label{fig:outflow}
\end{figure} 

\begin{figure}
   \epsscale{0.5}
   \plotone{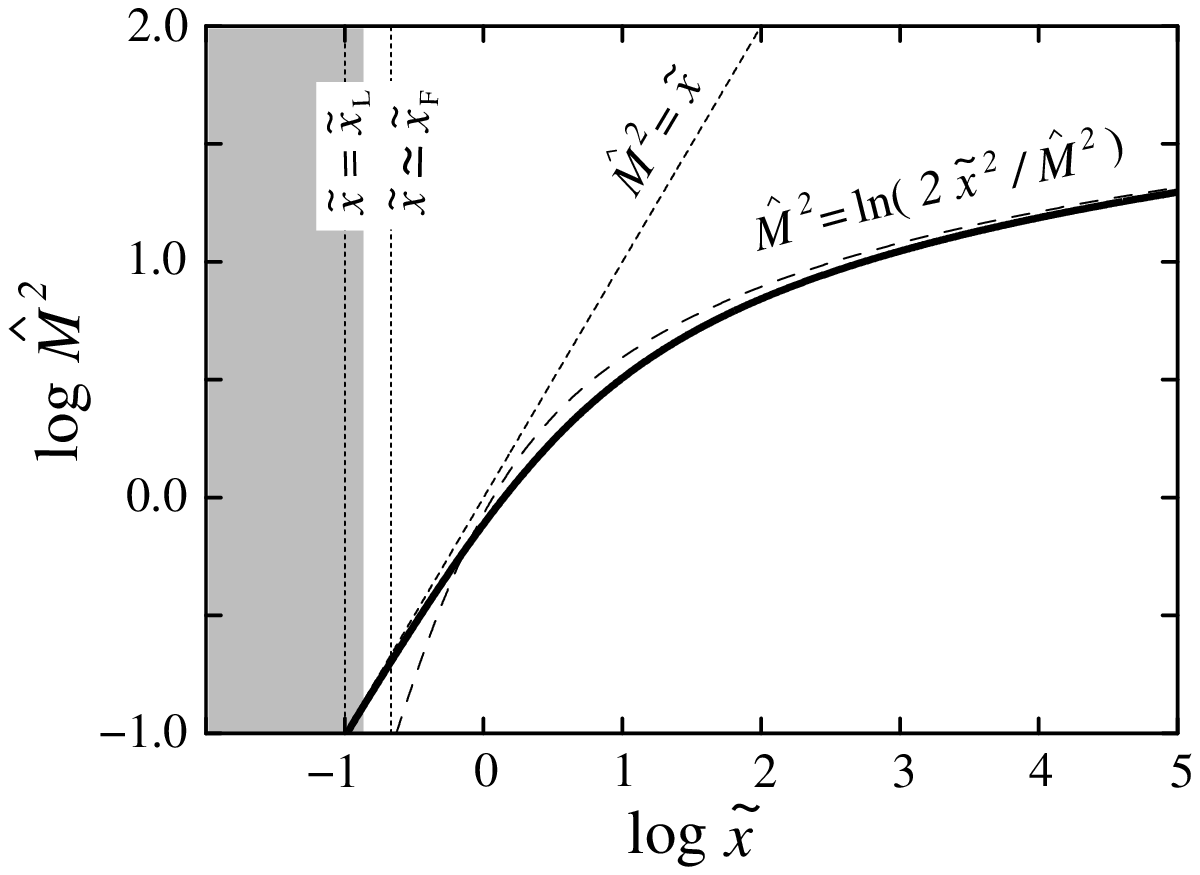} 
   \vspace{5mm}
   \plotone{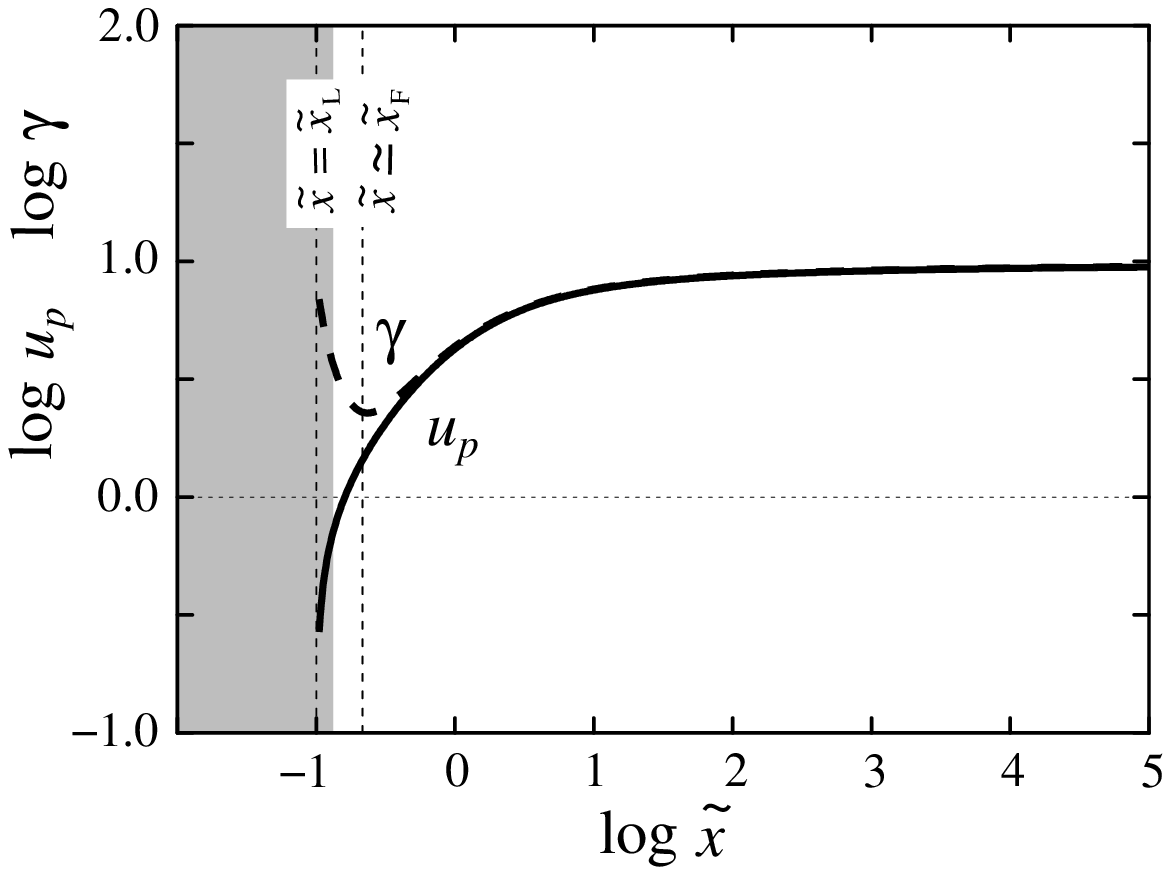} 
   \vspace{5mm}
   \plotone{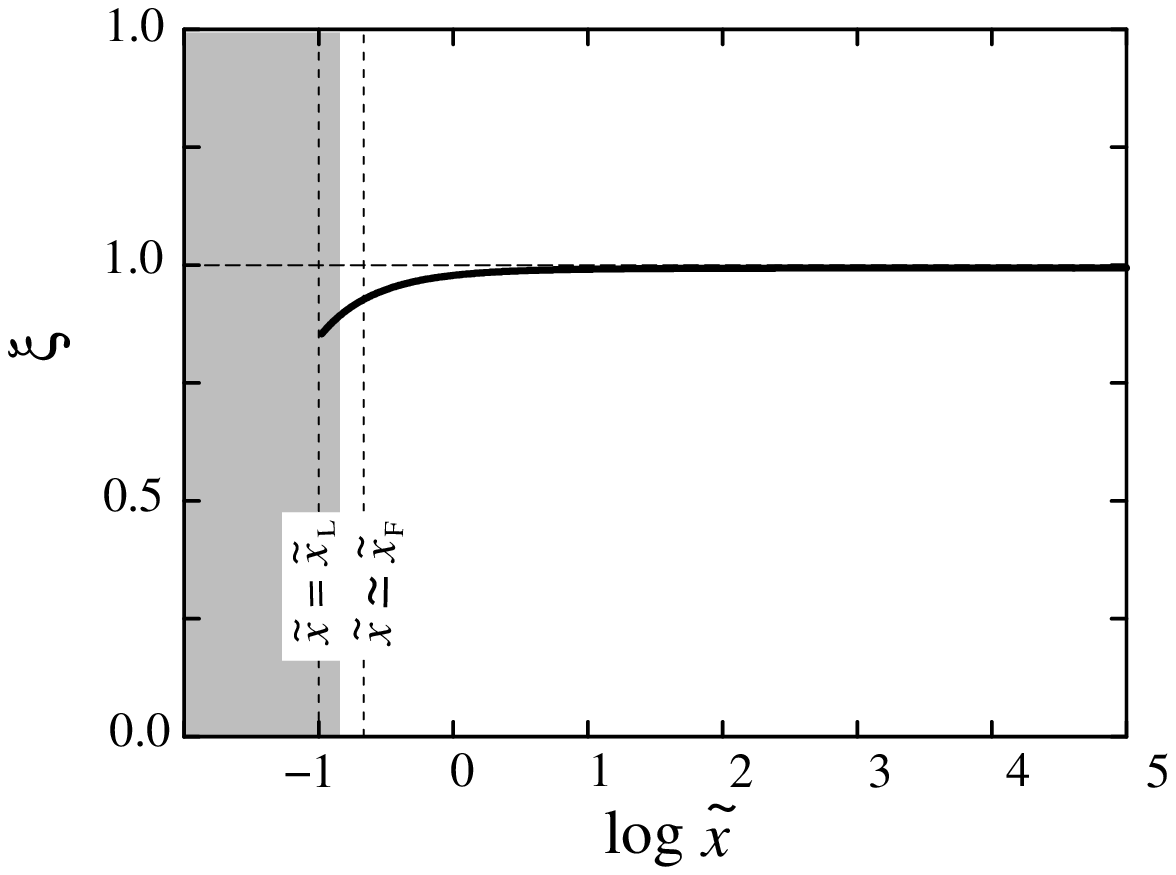} 
\caption{ 
 A solution of the approximated Grad-Shafranov equation given by
 equation (\ref{basic}) with $E=10.0$. (a) The Alfv\'{e}n Mach number 
 with two limiting curves ($\mm^2\simeq \y$ for $x_{\rm L}\ll \y \ll 1$ 
 and $\mm^2\simeq \ln(2\y^2/\mm^2)$ for $\y\gg 1$), (b) the poloidal
 velocity $u_p(\y)$ (solid) and the Lorentz factor $\gamma(\y)$
 (dashed), and (c) the ratio of the poloidal electric to toroidal
 electric magnetic field amplitude $\xi(\y)$.     
 This solution is valid for $\y > \y_{\rm L}=1/E$. 
        } 
\label{fig:gs-mm}
\end{figure} 

\begin{figure}
   \epsscale{0.75}
   \plotone{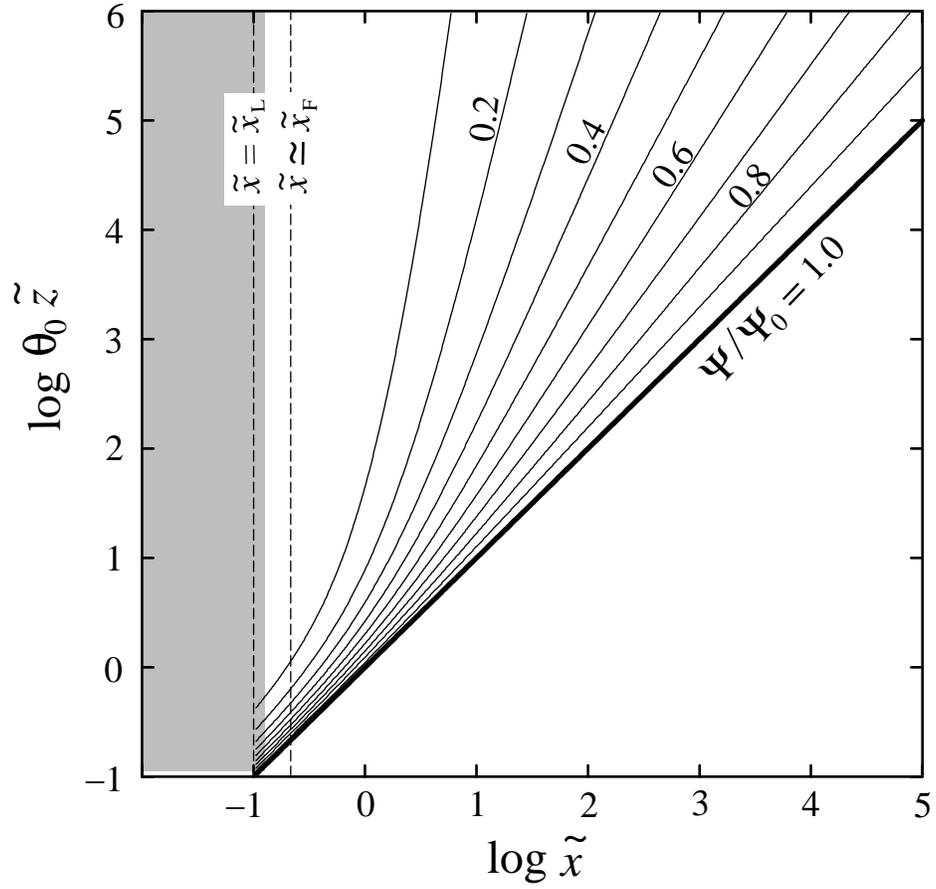} 
\caption{ 
 The magnetic field configuration corresponding to the solution given by
 Fig.\ref{fig:gs-mm}. It is assumed that the jet flow is confined within
 a cone of small opening angle $\theta_0$ by an external pressure. For
 the inner field lined of $\Psi<\Psi_0$, a paraboloidal collimation is
 obtained.   
        } 
\label{fig:z-x}
\end{figure} 

\begin{figure}
   \epsscale{0.7}
   \plotone{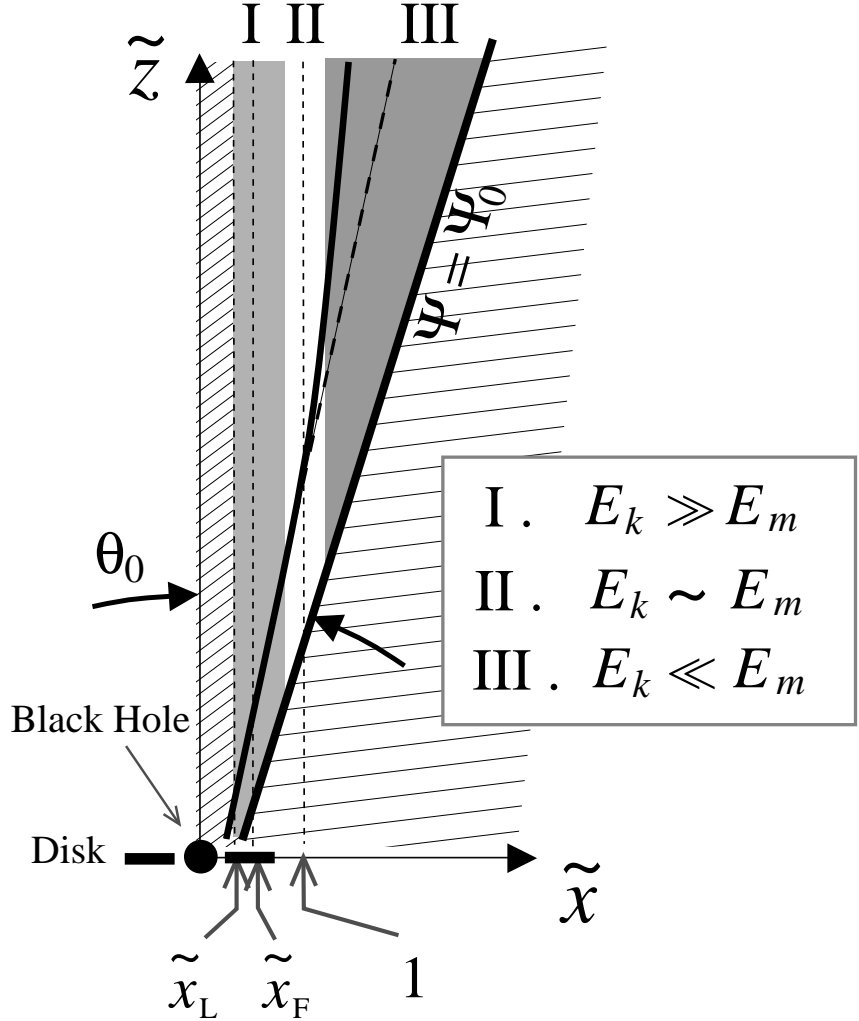} 
\caption{ 
 Schematic picture of the jet solution given by Fig.\ref{fig:z-x}. 
 The term $\theta_0$ is the opening angle of the last magnetic flux 
 surface.  The initial magnetically dominated outflow $E_{m}\gg E_{k}$ 
 accelerates along an almost radial field line at the intermediate 
 region  $\y_{\rm L} < \y < 1$ (region I), and then a rough equitation
 $E_{k}\sim E_{m}$ is realized around the radius $\tilde{x}\sim 1$ far
 beyond the light cylinder (region II). Subsequent logarithmic increase
 of $E_k$ goes on at large radii $\tilde x\gg 1$ (region III) along a
 collimating field line.   
        } 
\label{fig:jet}
\end{figure} 

\end{document}